\documentclass[twocolumn,prb,superscriptaddress,longbibliography]{revtex4-2}

\usepackage{amsmath,amsfonts,amssymb}
\usepackage{graphicx,color}
\usepackage{hyperref}
\usepackage{epstopdf}
\usepackage{float}
\usepackage{multirow}
\usepackage{hhline}
\usepackage{ulem}
\usepackage{mathtools}
\usepackage[caption=false]{subfig}
\usepackage{soul}

\newcommand{\ket}[1]{\left|#1\right\rangle}

\makeatletter

\@ifundefined{textcolor}{}
{%
 \definecolor{BLACK}{gray}{0}
 \definecolor{WHITE}{gray}{1}
 \definecolor{RED}{rgb}{1,0,0}
 \definecolor{GREEN}{rgb}{0,1,0}
 \definecolor{BLUE}{rgb}{0,0,1}
 \definecolor{CYAN}{cmyk}{1,0,0,0}
 \definecolor{MAGENTA}{cmyk}{0,1,0,0}
 \definecolor{YELLOW}{cmyk}{0,0,1,0}
}

\makeatother

\begin{document}

\title{Realization of two-qutrit quantum algorithms on a programmable superconducting processor}

\author{Tanay Roy$^\dag$}
\email{Present address: Superconducting Quantum Materials and Systems Center, Fermi National Accelerator Laboratory (FNAL), Batavia, IL 60510, USA}
\affiliation{James Franck Institute, University of Chicago, Chicago, Illinois 60637, USA}
\affiliation{Department of Physics, University of Chicago, Chicago, Illinois 60637, USA}
\altaffiliation{These authors contributed equally to this work}

\author{Ziqian Li$^\dag$}
\affiliation{James Franck Institute, University of Chicago, Chicago, Illinois 60637, USA}
\affiliation{Department of Physics, University of Chicago, Chicago, Illinois 60637, USA}
\altaffiliation{These authors contributed equally to this work}

\author{Eliot Kapit}
\affiliation{Department of Physics, Colorado School of Mines, Golden, CO 80401}

\author{David I. Schuster}
\affiliation{James Franck Institute, University of Chicago, Chicago, Illinois 60637, USA}
\affiliation{Department of Physics, University of Chicago, Chicago, Illinois 60637, USA}
\affiliation{Pritzker School of Molecular Engineering, University of Chicago, Chicago, Illinois 60637, USA}

\date{\today}

\begin{abstract}

Processing quantum information using quantum three-level systems or qutrits as the fundamental unit is an alternative to contemporary qubit-based architectures with the potential to provide significant computational advantages. We demonstrate a fully programmable two-qutrit quantum processor by utilizing the third energy eigenstates of two transmons. We develop a parametric coupler to achieve excellent connectivity in the nine-dimensional Hilbert space enabling efficient implementations of two-qutrit gates. We characterize our processor by realizing several algorithms like Deutsch-Jozsa, Bernstein-Vazirani, and Grover's search. Our efficient ancilla-free protocols allow us to show that two stages of Grover's amplification can improve the success rates of an unstructured search with quantum advantage. Our results pave the way for building fully programmable ternary quantum processors using transmons as building blocks for a universal quantum computer.

\end{abstract}

\maketitle

\section{Introduction}
Noisy intermediate-scale quantum (NISQ) computers~\cite{preskill2018nisq} are making rapid progress toward practical applications with demonstrated supremacy over classical algorithms for specific problems~\cite{Google2019supremacy, Pan2021advantage}. However, a majority of those processors utilize quantum two-level systems. Processing information using $d$-dimensional quantum systems or qudits can boost performance through access to a larger computational space and with fewer entangling gates for certain algorithms~\cite{Bullock2005qudit, Ralph2007qudit_Toffoli, Lanyon2009qudit, Gedik2015qudit, Nikolaeva2021qudit}. In the family of qudits, quantum three-level systems or qutrits are the closest members to qubits and provide the immediate opportunity to explore possibilities beyond two levels as basic units.

Theoretical studies show that qutrit-based processors can be beneficial for quantum error correction by providing compact logical encoding to protect against erasure~\cite{Kapit2016VSLQ, Muralidharan2017qec} and ternary errors~\cite{Majumdar2018qec}, enhance fault tolerance~\cite{Campbell2014fault_tolerant}, and enable magic state distillation~\cite{Campbell2012magic_state}. Other proposals claim improved implementations of quantum algorithms~\cite{Gokhale2019qutrit}, simulations~\cite{Gustafson2022qutrit}, cryptography~\cite{Bechmann2000qutrit_crypto, Bru2002qutrit_crypto}, and communication~\cite{Vaziri2002qutrit_communication} using qutrits. Due to these advantages, several platforms including photonic circuits~\cite{Luo2019qutrit_teleport, Hu2020qutrit_teleport}, trapped ions~\cite{Ringbauer2021qudit_ion, Hrmo2022qudit_ion}, and superconducting circuits~\cite{Siddiqi2021qutrit_RB, Siddiqi2022qutrit_CZ, Luo2022qutrit_CZ} have started to explore qutrits as computational units. In the superconducting circuits platform~\cite{cqed_review2021}, the transmon~\cite{koch2007transmon} is a natural choice to be utilized as a qutrit due to its fast gates~\cite{Siddiqi2021qutrit_RB, Luo2022qutrit_CZ}, long coherence times~\cite{Houck2021transmon_longlived, Wang2022transmon_longlived}, weak anharmonicity, and easy measurement~\cite{Wallraff2010qutrit_tomo}. While a significant effort is being made in realizing efficient two-qutrit gates ~\cite{Siddiqi2021qutrit_RB, Siddiqi2022qutrit_CZ, Luo2022qutrit_CZ} with demonstrations of solving specific problems~\cite{Siddiqi2021scrambling}, implementing multiple generic quantum algorithms on a single processor has remained a challenge in the circuit QED architecture~\cite{cqed_review2021} due to limitations arising from decoherence, slow, static inter-qutrit interaction or native entangling operations restricted to a smaller subspace.

In this work, we present a versatile two-qutrit processor with excellent connectivity between different states of the full Hilbert space facilitating a rich set of native entangling gates. The connectivity is achieved through multiple beam-splitting and two-photon squeezing-like primitive operations enabled by a linear parametric coupler~\cite{Yao2017stabilization, Ranzani2022stabilization}. We demonstrate two-qutrit versions of Deutsch-Jozsa~\cite{DJ_algo, Wang2020qudit}, Bernstein-Vazirani~\cite{BV_algo, Wang2020qudit}, and Grover's search~\cite{Grover_algo} algorithms without using any ancilla~\cite{Tanay2020Trimon2}. Deutsch-Jozsa and Bernstein-Vazirani algorithms provide exponential and linear speed-ups respectively over corresponding classical algorithms, whereas Grover's search provides a quadratic improvement. We perform two stages of Grover's amplification with success probabilities significantly larger than classically achievable values. To our knowledge, ours is the \textit{first successful demonstration} of a qutrit-based Grover's search across any quantum computing platform.

\section{Device description}

\begin{figure}[t]
\centering\includegraphics[width=\columnwidth]{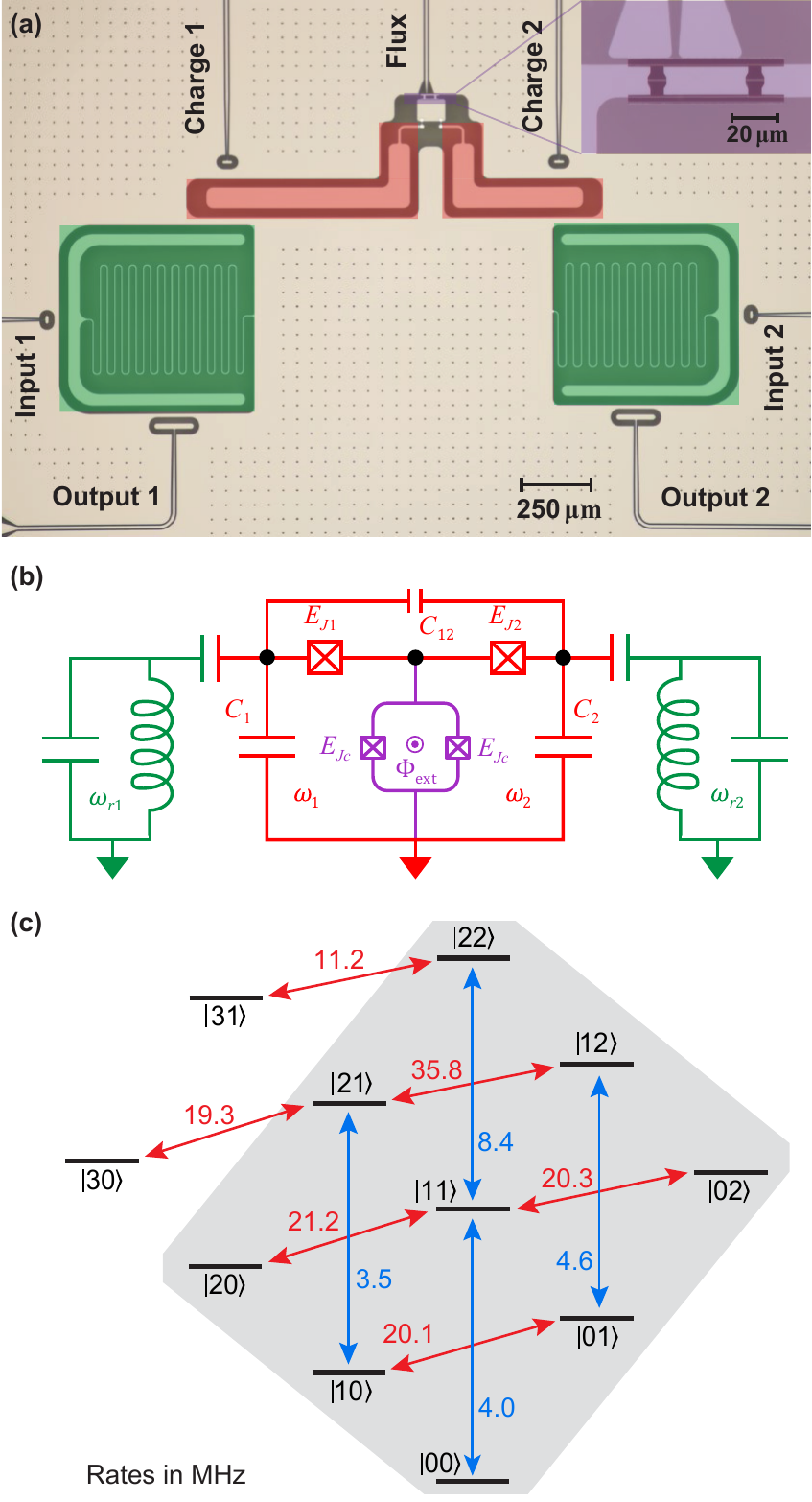}
\caption{The two-qutrit device. (a) False colored optical micrograph of the device. Two transmons (red) are inductively connected through a SQUID loop (purple, inset is an SEM image) that enables inter-qutrit parametric coupling. The transmons are capacitively coupled to linear resonators (green) for individual readout. Single-qutrit and readout pulses are sent through the input lines. (b) Simplified circuit diagram of the device. (c) Energy level diagram with various sideband interaction strengths (in MHz). The computational subspace is highlighted with a gray background and the other states are used as ancillary levels to implement controlled-phase gates. }
\label{fig:Device}
\end{figure}

Our two-qutrit processor, shown in Fig.~\ref{fig:Device}(a), is comprised of two transmons~\cite{koch2007transmon} ($Q_1$ and $Q_2$), each having its readout resonator. Parametric coupling~\cite{Yao2017stabilization} between the two transmons is realized by grounding the transmon junctions~\cite{Ranzani2022stabilization} through a superconducting quantum interference device (SQUID). The SQUID acts as a tunable inductor when an external DC magnetic field $\Phi_{\rm ext}$ is threaded into the loop. The transmon pads are also designed to provide a static capacitive coupling with a strength larger than the inductive coupling at zero external flux. By applying a finite $\Phi_{\rm ext}$ one can, thus, nullify the competing capacitive and inductive energies to minimize the cross-Kerr coupling. The fast-flux line is used to enable various inter-qutrit interactions by modulating the SQUID at appropriate radio frequencies, whereas the charge lines can be utilized to drive qutrit-resonator sidebands. Individual manipulation of the qutrits are done by sending appropriate microwave pulses to the resonators, and the states of the qutrits are determined through transmission measurements. We keep large Josephson-to-charging energy ratios ($\gtrsim$ 90) to maintain long dephasing times for the second excited states $(\ket{2})$ for both transmons~\cite{koch2007transmon}. Figure~\ref{fig:Device}(b) shows the equivalent circuit diagram.

\section{Hamiltonian and operating point}
The system can be described by the following Hamiltonian when no parametric drives are present
\begin{equation}
    \dfrac{H}{\hbar} = \sum_{k=1}^2\omega_k \hat{n}_{k} + \dfrac{\alpha_k}{2} \hat{n}_{k}(\hat{n}_k-1) + 
    \sum_{j,k\ge1}J_{jk} (\hat{n}_1)^j (\hat{n}_2)^k,
\end{equation}
where $\omega_k$ is $\ket{0}\leftrightarrow \ket{1}$ transition frequency, $\alpha_k$ represents anharmonicity, and $\hat{n}_k = a^\dagger_k a_k$ is the photon number operator with 
$a^\dagger_k (a_k)$ being the creation (annihilation) operator for the $k$-th qutrit. At $\Phi_{\rm ext}=0$, only second order parametric processes are allowed, whereas biasing the SQUID loop sufficiently away from the sweet-spot enables first order processes. The cross-Kerr couplings $J_{jk}$ also tune with $\Phi_{\rm ext}$ and we choose a bias flux $\Phi_{\rm min}$ so that the leading order dispersive shift $J_{11}$ is minimized. We treat $\Phi_{\rm min}$ as the coupler's `off' point as the higher order cross-Kerr shifts $J_{12}$ and $J_{21}$ are very close to minimum at this bias (see Appendix~\ref{app:Hamiltonian} for details) ensuring reduced crosstalk during idling periods or gate operations. The value of $\Phi_{\rm min}$ is experimentally determined by performing Ramsey-fringe experiments on $Q_1$ conditioned on $Q_2$ being in the ground or first excited state and minimizing the difference in the oscillation frequencies with respect to $\Phi_{\rm ext}$. 

We activate qutrit-qutrit (QQ) red and blue sidebands corresponding to photon-exchange (beam splitting) $\ket{m,n}\leftrightarrow\ket{m-1,n+1}$ and two-photon pumping (squeezing) $\ket{m,n}\leftrightarrow\ket{m+1,n+1}$ interactions respectively by modulating the SQUID at corresponding transition frequencies (see Fig.~\ref{fig:Device}(c)). Here the state of the system $\ket{m,n}$ indicates the qutrits being in the energy eigenstates $\ket{m}$ and $\ket{n}$ respectively. The SQUID (coupler) mode is designed to have a frequency much larger than any relevant radio-frequency (RF) drives during the experiment at the operating point and is never populated. The blue sideband frequencies (around 6.5 - 6.9 GHz) are usually one order of magnitude higher than the red sidebands (around 0.2 - 0.7 GHz). 
For the current geometry, stray capacitive coupling of the flux line to the SQUID~\cite{Ranzani2022stabilization} limits the blue sideband rates due to frequency-enhanced unwanted interactions.

\section{Gate operations}
Single-qutrit operations are performed by applying separate microwave pulses at $\ket{0}\leftrightarrow \ket{1}$ and $\ket{1}\leftrightarrow \ket{2}$ transition frequencies of both qutrits through the readout resonators. We use Gaussian-edge rectangular pulses having $2\sigma$ edge lengths. Single-qutrit drives use roughly 5 MHz of Rabi rates, whereas phase gates are realized by simply advancing the phases of the appropriate subsequent pulses. Consequently, the virtual phase gates take no time and are near-perfect. For the quantum algorithms we use qutrit-Hadamard $H$, bit-shift $X$ (extension of the qubit's NOT gate), and qutrit-phase $Z$ gates having the following matrix representations
\begin{equation}
    H = \dfrac{1}{\sqrt{3}}
    \begin{bmatrix}
    1 & 1 & 1 \\
    1 & \omega & \omega^2 \\
    1 & \omega^2 & \omega
    \end{bmatrix}, 
    X = 
    \begin{bmatrix}
    0 & 0 & 1 \\
    1 & 0 & 0 \\
    0 & 1 & 0
    \end{bmatrix}, 
    Z = 
    \begin{bmatrix}
    1 & 0 & 0 \\
    0 & \omega & 0 \\
    0 & 0 & \omega^2
    \end{bmatrix}, 
\end{equation}
where $\omega=e^{2\pi i/3}$, the cube-root of unity. Implementation of $H,X,$ and $Z$ gates require the application of three, two, and zero physical pulses respectively (see Appendix~\ref{app:single_gate}).

\begin{figure}[t]
    \centering
    \includegraphics[width=1.0\columnwidth]{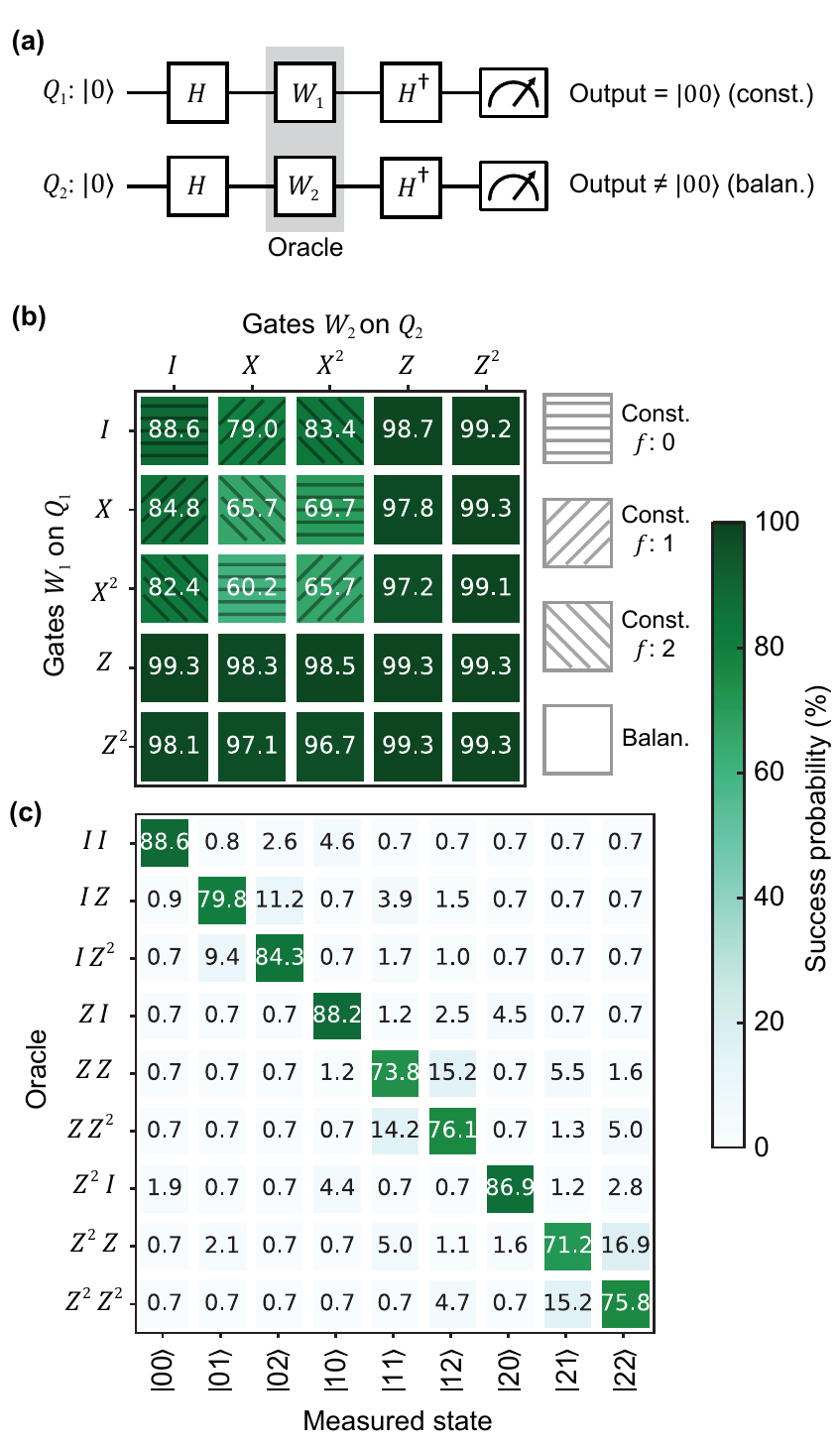}
    \caption{Deutsch-Jozsa and Bernstein-Vazirani algorithms. (a) Quantum circuit for the algorithms. In DJ algorithm, gates $W_{1}$, $W_{2} \in{\{I,X,X^2,Z,Z^2\}}$ are applied to implement a constant or a balanced oracle. The final output state being in $\ket{00}$ or non $\ket{00}$ distinguishes the two cases. (b) Experimental results for DJ algorithm. The rows and columns represent gates applied to $Q_{1}$ and $Q_{2}$ respectively. The average SPs are 75.5(3)\% and 98.5(1)\% for the constant (hatched boxes) and balanced (plain boxes) oracles respectively beating the classical rate of 50\%. (c) Experimental results for BV algorithm. Each row corresponds to a specific oracle with the mapping $\{I, Z, Z^{2}\} \rightarrow \{ 0,1,2 \}$. The diagonal terms show the SPs for all nine strings mapped to the basis states with an average of 78.3(3)\%, which is much larger than the classical SP of 33.3\%.}
    \label{fig:DJ}
\end{figure}

We can implement arbitrary two-qutrit unitaries by combining single-qutrit rotations along with generalized controlled-phase (CPhase) gates $C_p(\theta,\ket{mn}) = \mathcal{I} - (1-e^{i\theta})\ket{mn}$ representing an accumulation of $\theta$ phase on the eigenstate $\ket{mn}$, where $\mathcal{I}$ is the nine-dimensional identity matrix. We realize $C_p(\pi,\ket{22})$ gate by applying a 94~ns long $2\pi$-pulse at the red sideband $\ket{22} \leftrightarrow \ket{31}$. Similarly, a 56~ns $2\pi$-pulse at the red sideband $\ket{21} \leftrightarrow \ket{30}$ enables a $C_p(\pi,\ket{21})$ gate (see Fig.~\ref{fig:Device}(c)). We sandwich these $C_\theta$ gates between single-qutrit $\pi$ gates to achieve phase flip on any computational basis (see Appendix~\ref{app:CPhase}). While we have access to a native $C_p(\pi,\ket{12})$ gate, we avoid that due to lower fidelity. Note that the bosonic enhancement of the coupling to an external field makes $C_p(\pi,\ket{22})$ twice as fast compared to $C_p(\pi,\ket{11})$ when restricting to the qubit subspace. Next, we implement three two-qutrit quantum algorithms on our processor.

\section{Deutsch-Jozsa algorithm}

The Deutsch-Jozsa (DJ) algorithm\cite{DJ_algo} is one of the earliest quantum algorithms showing an exponential advantage over any classical algorithm. The original DJ algorithm applies to qubits, and here we extend it to qutrits~\cite{Wang2020qudit} for our quantum processor. For an $n$-qutrit system, the task of the DJ algorithm is to distinguish a function $f:\{0,1,2\}^{n}\rightarrow\{0,1,2\}$, which takes $n$-trits as an input and outputs one trit, between two cases, a balanced or a constant function. The constant function always results in the same output (0, 1, or 2) independent of the input, whereas the balanced function outputs each of the three possibilities for exactly one-third of the possible inputs. Note that implementing the different test functions is often termed as the \textit{oracle}. A deterministic classical algorithm needs $3^{n-1}+1$ queries (with at least two queries in the best case) to distinguish the two cases, whereas DJ algorithm needs only one and hence provides the exponential speed-up. 

Figure.~\ref{fig:DJ}(a) depicts our (\textit{ancilla-free}) circuit implementation of the DJ algorithm. Two Hadamard gates are simultaneously applied to both qutrits initialized to $\ket{0}$ (ground state) to prepare the state $\frac{1}{3}\left(\ket{0}+\ket{1}+\ket{2}\right)^{\otimes 2}$. The oracles (gray gates in Fig.~\ref{fig:DJ}(a)) are implemented by applying gates $(W)$ to the qutrits chosen from the set $\mathcal{S} = \{I, X, X^{2}, Z, Z^{2}\}$. For the constant case, $W_{1}$ and $W_{2}$ are picked up from the subset $\mathcal{S}_c=\{I, X, X^{2}\}$, and the total number of $X$ gates modulo $3$ specifies the constant output. For example, the gate $X\otimes X^2 $ would implement constant 0, whereas $I\otimes X^2$ would be the case of constant 2. A balanced oracle can be realized by choosing any combination of elements from the set $\mathcal{S}$ except for those cases that result in a constant function. 16 different kinds of balanced functions (see Appendix~\ref{app:balanced} for mapping to various addition functions) are implemented in our experiment as shown in Fig.~\ref{fig:DJ}(a). Finally, two $H^\dagger$ gates are applied before simultaneous readout. A final measured state of $\ket{00}$ indicates a constant function, whereas any other output implies a balanced function. The theoretical success probability (SP) for each case is 100\%, and the experimental results are summarized in Fig.~\ref{fig:DJ}(b) where the hatched (unhatched) boxes represent constant (balanced) cases. The average SPs for the three constant cases with outputs $\{0,1,2\}$ are separately $72.8(3)\%$, $76.5(3)\%$ and $77.2(3)\%$, marked with horizontal, left and right hatching. The numbers in parentheses represent standard error of mean obtained after 20,000 repetitions of each oracle. All experimental data are corrected for measurement error (see Appendix~\ref{app:error_mitigation}). For the 16 balanced cases, the average SP is $98.5(1)\%$. The SPs for all cases are well above the classical case, which would be 50\% after a single query.

\section{Bernstein-Vazirani algorithm}

The Bernstein-Vazirani (BV) algorithm\cite{BV_algo} for qutrits can be restated as follows: given an oracle $f\left(\boldsymbol{x,s}\right)\equiv\sum_{j=1}^{n} x_{j}s_{j} \pmod{3}$ that performs inner product between two strings of ternaries followed by modulo 3, the goal is to determine the unknown string $\boldsymbol{s}=\{s_1,s_2,\ldots,s_n\}$ where the user has control over the input string $\boldsymbol{x}$. The most efficient classical algorithm will need $n$ oracle queries to find all digits of $\boldsymbol{s}$. BV algorithm, on the contrary, needs only one query and the quantum circuit is identical to the case of DJ as shown in Fig.~\ref{fig:DJ}(a). Oracles representing 9 different strings for our two-qutrit system are implemented by choosing gates ($W$) from the set $\{I, Z, Z^{2}\}^{\otimes 2}$ with the mapping $\{I, Z, Z^{2}\} \rightarrow \{ 0,1,2 \}$. The final state of the system after measurement directly reveals the unknown string with 100\% theoretical success rate. Fig.~\ref{fig:DJ}(c) tabulates the experimental results where the vertical axis represents gates applied to the qutrits corresponding to different unknown strings, and the horizontal axis shows the measured probability for each state. The diagonal entries indicate individual SPs for each input string mapped to the final state. The average SP for all $9$ cases is $78.3(3)\%$, which is far above the classical SP of 33.3\% after one query.

\begin{figure}[t]
    \centering
    \includegraphics[width=1.0\columnwidth]{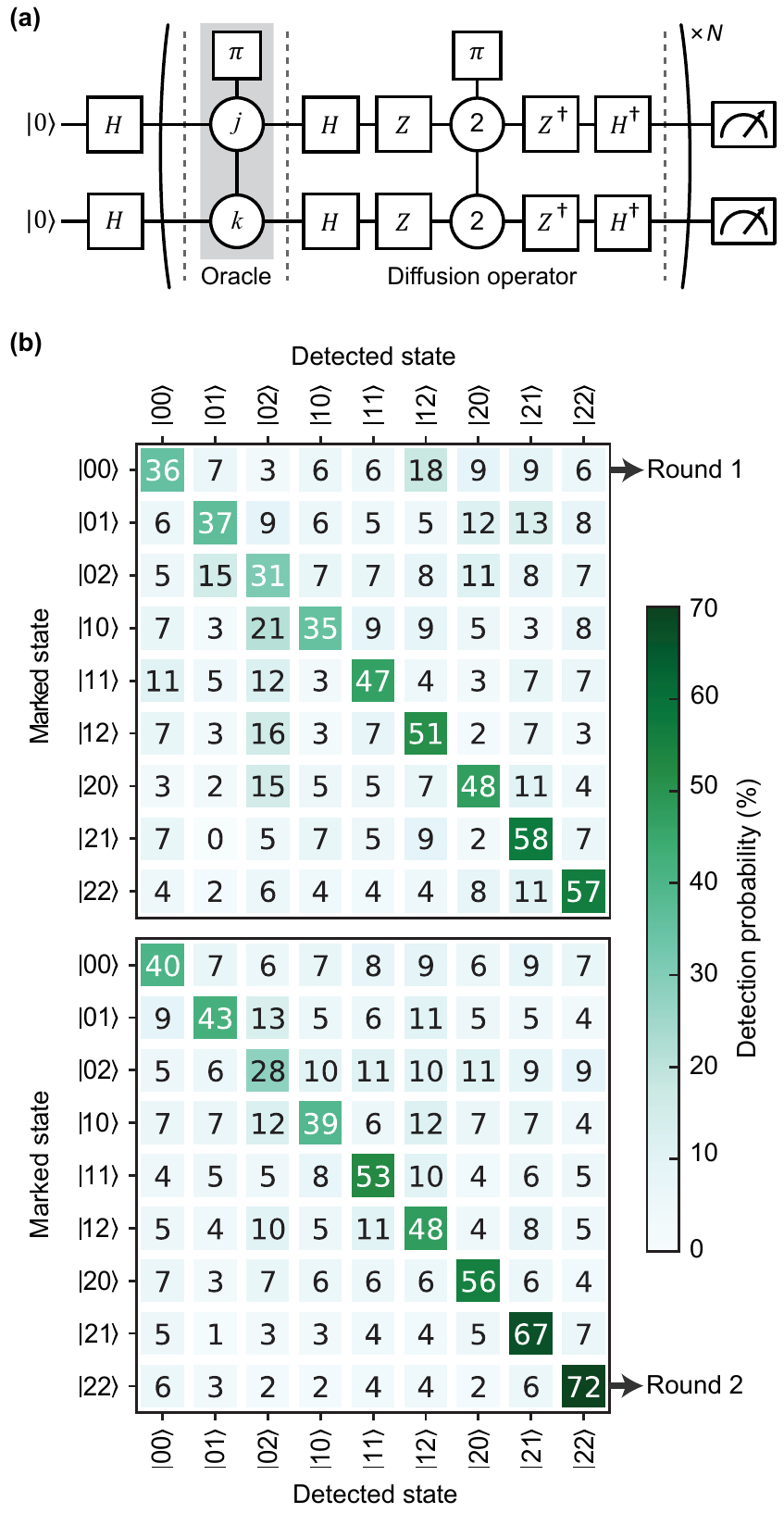}
    \caption{Grover's search algorithm for two-qutrits. (a) Quantum circuit. The oracles are implemented by CPhase gates $C_{\theta}\left(\pi,\ket{jk}\right)$. The diffusion operator amplifies the detection probability of the marked state. (b) Experimental results. Detection probabilities (corrected for measurement error) after one (top panel) and two (bottom panel) rounds of amplitude amplification are obtained with 20,000 averages. All individual success rates are far beyond the corresponding classical SPs of 11.1\% and 22.2\%.}
    \label{fig:Grover}
\end{figure}

\section{Grover's Search}

Grover's algorithm~\cite{Grover_algo} provides a quadratic speed-up for searching an unstructured database. For a database of size $\mathcal{N}$, the algorithm can find the unique input that satisfies a certain condition using $O\left(\sqrt{\mathcal{N}}\right)$ search queries, while a classical algorithm requires on average ${\mathcal{N}}/{2}$ repetitions. Several groups have recently realized Grover's search on qubit-based platforms~\cite{Grover-trapped-ion, Tanay2020Trimon2, Grover234_ibm, Grover345_ibm}. For the two-qutrit case with $\mathcal{N}=3^2=9$, the classical SPs with one and two rounds of search are $\frac{1}{9}=11.1\%$ and $\frac{1}{9}+\frac{8}{9}\cdot\frac{1}{8}=22.2\%$ respectively. The corresponding theoretical SPs for the original Grover's search are $72.6\%$ and $98.4\%$, and can also be modified to achieve determinism~\cite{Tanay2022Grover, Long2001certain_Grover}.

Our ancilla-free version of the two-qutrit Grover's search is illustrated in Fig.~\ref{fig:Grover}(a). It has four stages: initialization, oracle implementation, amplitude amplification, and measurement. Starting from the ground state, we apply Hadamard gates on both qutrits to initialize the system to the equal superposition state $\frac{1}{3}\left(\ket{0}+\ket{1}+\ket{2}\right)^{\otimes 2}$. A controlled-phase gate $C_{\theta}\left(\pi,\ket{jk}\right)$, that flips the phase of the target state $\ket{jk}$, is used to realize the oracle. Due to the structure of the Hilbert space (see Fig~\ref{fig:Device}(c)), we have access to two high-fidelity CPhase gates $C_{\theta}\left(\pi,\ket{22}\right)$ and $C_{\theta}\left(\pi,\ket{21}\right)$ and all other CPhase gates are realized in conjunction with single-qutrit rotations (see Appendix~\ref{app:CPhase}). Amplitude amplification of the marked state happens through the Grover's diffusion or reflection unitary, which is constructed using a phase flip of the $\ket{22}$ state sandwiched between Hadamard and $Z$ gates. Here, we utilized the decomposition $C_{\theta}\left(\pi,\ket{00}\right) = (ZZ) \otimes C_{\theta}\left(\pi,\ket{22}\right) \otimes (Z^\dagger Z^\dagger)$. Simultaneous measurements on the qutrits are performed after one and two iterations of Grover's search for each target state. During a similar search using three qubits ($\mathcal{N}=8$), each oracle (and amplification) step requires eight CNOT gates (for a linear chain)~\cite{Grover345_ibm}, resulting in an \textit{eight-fold rise} in entangling operations compared to our efficient two-qutrit implementation.

Figure~\ref{fig:Grover}(b) shows the experimentally obtained detection probabilities for the 9 different marked states after single (top panel) and double (bottom panel) rounds of the Grover's iteration.  Each row represents a probability distribution acquired with 20,000 repetitions and after correcting for measurement error. The diagonal terms represent successful detection rates with an average SP of $44.4(3)\%$ after the first round, which increases to $49.6(3)\%$ with the second iteration. The performance degradation of target states closer to $\ket{00}$ is caused by the less-efficient implementations of the corresponding oracles, where more single-qutrit rotations are required for the CPhase gate decomposition (see Appendix~\ref{app:CPhase}). As promised by the algorithm, experimental detection probabilities for the individual correct states increased after the second iteration for all cases (except for $\ket{02}$ and $\ket{12}$, which we attribute to the lower lifetime of $Q_2$'s $\ket{2}$ level). The experimental outcome also has a good agreement with our Master equation simulation as described in Appendix~\ref{app:error} indicating dephasing limited performance. The average experimental SPs beat the classical rates of 11.1\% and 22.2\% for the two rounds by more than a factor of 2, with clear improvement in performance after the second iteration.

\section{Conclusion}

We demonstrate a fully programmable two-qutrit superconducting processor based on transmons. We utilize a linear coupler to obtain excellent connectivity in the underlying Hilbert space through parametrically activated fast sidebands. The success probabilities of all the algorithms implemented, namely Deutsch-Jozsa, Bernstein-Vazirani, and Grover's search, are significantly higher than corresponding classical rates. Notably, we achieve improved success rates in finding the correct answer after performing the second iteration of Grover's search. To our knowledge, this is the \textit{first experimental demonstration} of two-qutrit Grover's search across all quantum computing platforms. The performance of the processor is currently limited by low dephasing times, and we expect significant improvement with increased coherence~\cite{Houck2021transmon_longlived, Wang2022transmon_longlived}, pulse shaping~\cite{Filipp2021optimal_control}, geometrical optimization, and improved readout~~\cite{cqed_review2021}. Our results indicate that qutrit-based processors could be a promising candidate for building a large-scale quantum computer and we believe, will foster investigations in the field of ternary quantum information processing.

\section{Acknowledgements}
This work was supported by AFOSR Grant No. FA9550-19-1-0399 and ARO Grant No. W911NF-17-S-0001. Devices are fabricated in the Pritzker Nanofabrication Facility at the University of Chicago, which receives support from Soft and Hybrid Nanotechnology Experimental (SHyNE) Resource (NSF ECCS-1542205), a node of the National Science Foundation’s National Nanotechnology Coordinated Infrastructure. This work also made use of the shared facilities at the University of Chicago Materials Research Science and Engineering Center.

\appendix

\section{Single-qutrit gates}
\label{app:single_gate}

The native operations accessible for a single qutrit are
\begin{subequations}
\label{eq:single_qutrit}
\begin{align}
    R_{01}\left(\phi,\theta\right) &= 
    \begin{bmatrix}
               \cos{\frac{\theta}{2}} & -e^{-i\phi}\sin{\left(\theta/2\right)} & 0 \\
           e^{i\phi}\sin{\frac{\theta}{2}} & \cos{\frac{\theta}{2}} & 0 \\
           0 & 0 & 1
    \end{bmatrix},
    \\
    R_{12}\left(\phi,\theta\right) &= 
    \begin{bmatrix}
               1 & 0 & 0 \\
           0 & \cos{\frac{\theta}{2}} & -e^{-i\phi}\sin{\left(\theta/2\right)} \\
           0 & e^{i\phi}\sin{\frac{\theta}{2}} & \cos{\frac{\theta}{2}}
    \end{bmatrix},
    \\
    \Theta\left(x,y\right) &= 
    \begin{bmatrix}
               1 & 0 & 0 \\
               0 & e^{ix} & 0 \\
               0 & 0 & e^{i\left(x+y\right)}
    \end{bmatrix},
\end{align}
\end{subequations}
where, $R_{01}\left(\phi, \theta \right)$ and $R_{12}\left(\phi, \theta \right)$ are performed through sending drives at $\ket{0}\leftrightarrow \ket{1}$ and $\ket{1}\leftrightarrow \ket{2}$ transitions respectively with appropriate lengths and phases. However, the generic phase gate $\Theta\left(x,y\right)$ is implemented virtually as it requires no physical pulses. For each qutrit we record two phase parameters $\theta_{01}$ and $\theta_{12}$ corresponding to the pulses applied to the $\ket{0}\leftrightarrow \ket{1}$ and $\ket{1}\leftrightarrow \ket{2}$ transitions respectively. To apply the  $\Theta\left(x,y\right)$ gate, we advance both $\theta_{01}$ and $\theta_{12}$ by $x$ and $y$ for all the subsequent pulses. The single-qutrit $Z$ gate becomes a special case of the generic phase gate: $Z=\Theta\left(\frac{2\pi}{3},\frac{2\pi}{3}\right)$. Since, all phase updates are performed in software, the $Z$ gate (or a $\Theta\left(x,y\right)$ gate) has nearly $100\%$ fidelity. 

An arbitrary single-qutrit unitary can be constructed using combinations of Eqs.~\eqref{eq:single_qutrit}. For example, the $H$ gate is deconstructed as
\begin{multline}
    \label{eq:H_decompose}
    H = \frac{1}{\sqrt{3}}
    \begin{bmatrix}
               1 & 1 & 1 \\
               1 & e^{\frac{2\pi}{3} i} & e^{\frac{4\pi}{3} i} \\
               1 & e^{\frac{4\pi}{3} i} & e^{\frac{2\pi}{3} i} 
    \end{bmatrix}
    \\
    = R_{12}\left(0, \frac{\pi}{2}\right) \cdot R_{01}\left(0, \beta\right) \cdot \Theta\left(\pi, \frac{\pi}{2}\right) \cdot R_{12}\left(0, \frac{\pi}{2}\right) \cdot \Theta\left(0, \pi\right),
\end{multline}
with $\beta=2 \tan^{-1}(\sqrt 2)$. Similarly, the bit-shift gate is decomposed as
\begin{equation}
    X = \begin{bmatrix}
               0 & 0 & 1 \\
               1 & 0 & 0 \\
               0 & 1 & 0 
    \end{bmatrix} 
    = R_{01}(0,\pi) \cdot R_{12}(0,\pi).
\end{equation}

\begin{figure*}[t]
    \centering
    \includegraphics[width=\textwidth]{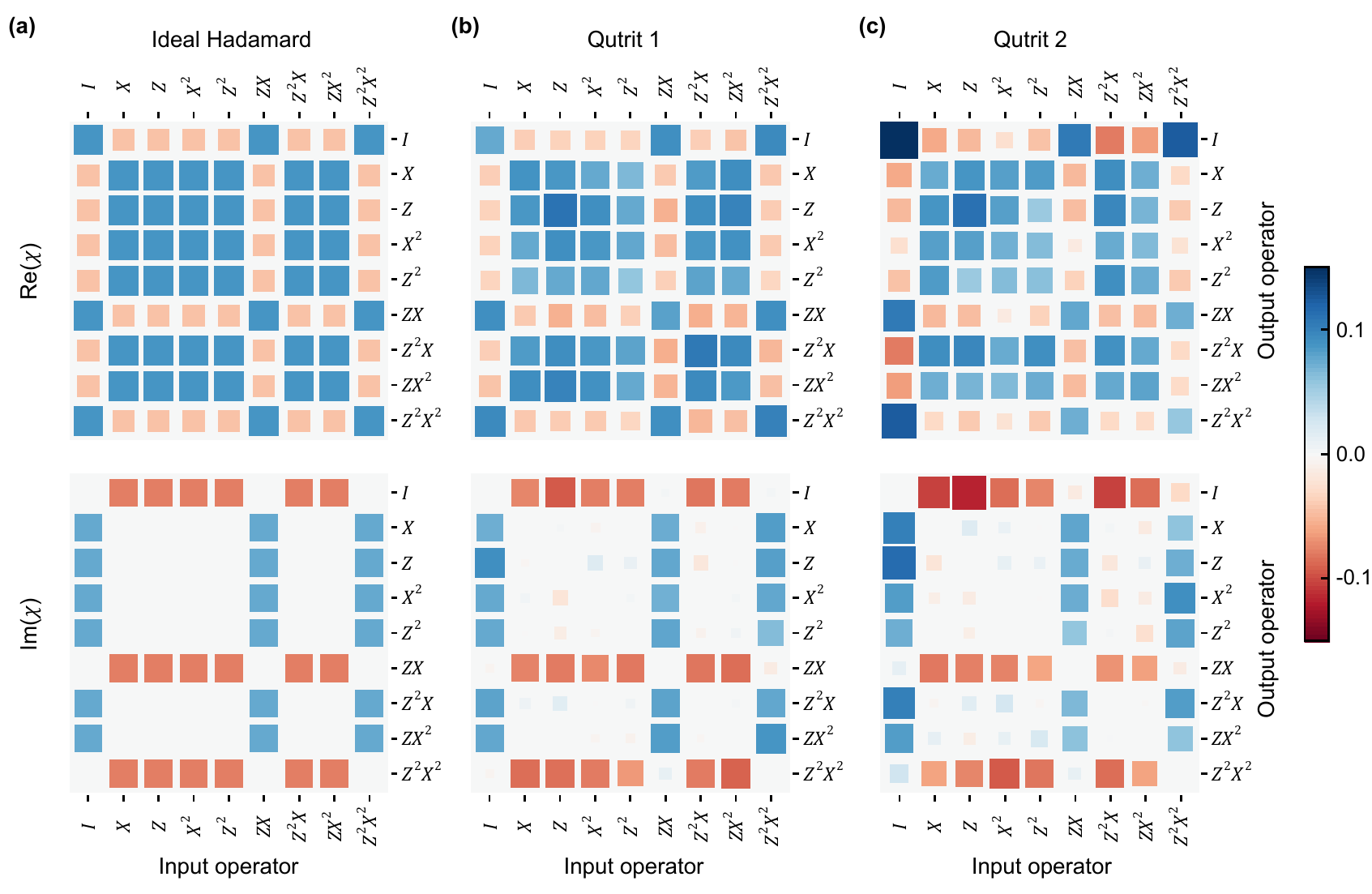}
    \caption{Process tomography for Hadamard gates. The top and bottom panels show real and imaginary components of the process matrix ($\chi$) for (a) an ideal case, (b) $Q_1$, and (c) $Q_2$.}
    \centering
    \label{fig:hadamard}
\end{figure*}

We obtain process fidelities of 98.96\% ($Q_1$) and 97.06\% ($Q_2$) for the $H$ gate as shown in Fig.~\ref{fig:hadamard}. The same for the $Z$ gate are 97.48\%  ($Q_1$) and 96.76\%  ($Q_2$). Even though the $Z$ gates should be nearly perfect, the process fidelities are limited by state preparation and measurement (SPAM) error.

\begin{table}[b]
\centering

\begin{tabular}{||c|c|c||}
    \hline
    Gate type & On $Q_{1}$ (ns) & On $Q_{2}$ (ns) \\
    \hline
    $\pi/2_{01}$ & 49.50 & 49.71 \\
    \hline
    $\pi/2_{12}$ & 41.27 & 44.15 \\
    \hline
    $\pi_{01}$ & 94.98 & 95.41 \\
    \hline
    $\pi_{12}$ & 78.52 & 84.28 \\
    \hline
    $Z$ & 0 & 0 \\
    \hline
    $H$ & 141.88 & 147.90 \\
    \hline
\end{tabular}
\caption{Total gate lengths for different single-qutrit operations. The $Z$ gates are implemented virtually.}
\label{tb:singlequbit}
\end{table}
All single-qutrit gates use rectangular pulses with Gaussian edges. The envelope shape $h\left(t\right)$ is defined in Eq.~\ref{eq:pulse} with $\sigma=2.5$ ns,
\begin{equation}
  h\left(t\right) =
    \begin{cases}
      A_{0}e^{\frac{-\left(t-t_{0}-2\sigma\right)^2}{2\sigma^2}} & \text{if $t_{0}<t<t_{0}+2\sigma$,}\\
      A_{0} & \text{if $t_{0}+2\sigma<t<t_{1}-2\sigma$,}\\
      A_{0}e^{\frac{-\left(t_{1}-2\sigma-t\right)^2}{2\sigma^2}} & \text{if $t_{1}-2\sigma<t<t_{1}$,}\\
      0 & \text{Other cases},\\
    \end{cases}       
\label{eq:pulse}
\end{equation}
where, $2\sigma$ is the Gaussian tail length, $A_{0}$ is the amplitude, and $t_1-t_0$ is the total pulse length.

\section{Device fabrication and parameters}
The device was fabricated on a 430~$\mu$m thick C-plane sapphire substrate annealed at $1200^o$C for 2 hours. A 200~nm thick film of Tantalum was sputtered at $800^o$C to form the ground plane. The large features (excluding the Josephson junctions) were made via optical lithography followed by wet-etching (dipped for 20 seconds) at wafer-scale using Transene Tantalum etchant 111. A 600~nm thick layer of AZ 1518 was used as the (positive) photoresist, and a Heidelberg MLA 150 Direct Writer was used for the photolithography. The junction mask was fabricated via electron-beam lithography using a Raith EBPG5000 Plus E-Beam Writer on a bi-layer resist (MMA-PMMA) comprising of MMA EL11 and 950PMMA A7. Both small (transmon) and large (coupler) Josephson junctions were made with the Dolan bridge technique. Those were subsequently evaporated in a Plassys electron-beam evaporator with double angle evaporation ($\pm 23^o$). The wafer was then diced into $7\times7$ mm chips, lifted-off, mounted on a printed circuit board, and subsequently wire-bonded. The coherence times and frequency parameters are listed in Table.~\ref{table:qutrit_coherence} and Table.~\ref{table:frequency}.

\begin{table}[b]
	\begin{tabular}{||c|c|c|c|c||}
	    \hline
	     Qutrit & $T_{1}^{01}(\mu$s) & $T_{1}^{12}(\mu$s) & $T_{2R}^{01}(\mu$s) & $T_{2R}^{12}(\mu$s) \\  \hline
	    $Q_{1}$ & $47.9$ & 21.7 & $4.5$ & $2.0$ \\  \hline
	    $Q_{2}$ & $35.1$ & 3.9  & $3.2$ & $2.4$ \\  \hline
	\end{tabular}
	\caption{Device coherence parameters. $T^{jk}_1$ and $T^{jk}_{2R}$ respectively represent the relaxation and Ramsey dephasing time constants for the $\ket{j} \leftrightarrow \ket{k}$ transition.}
	\label{table:qutrit_coherence}
\end{table}

\begin{table}[t]
	\begin{tabular}{||c|c|c||}
	    \hline
	     Parameter & Symbol & Value$/2\pi$ \\  \hline
	     $Q_{1}$ $\ket{0} \leftrightarrow \ket{1}$ transition & $\omega_{1}$ & $3.3494$ (GHz)  \\  \hline
	     $Q_{2}$ $\ket{0} \leftrightarrow \ket{1}$ transition & $\omega_{2}$ & $3.8310$ (GHz)  \\  \hline
	     $Q_{1}$ anharmonicity & $\alpha_{1}$ & $-115.2$ (MHz)  \\  \hline
	     $Q_{2}$ anharmonicity & $\alpha_{2}$ & $-159.8$ (MHz)  \\  \hline
	     Readout1 frequency & $\omega_{r1}$ & $4.9602$ (GHz)  \\  \hline
	     Readout2 frequency & $\omega_{r2}$ & $5.4225$ (GHz)  \\  \hline
	     $\left(E_{\ket{11}}-E_{\ket{01}}\right)-\left(E_{\ket{10}}-E_{\ket{00}}\right)$ & $ZZ$ & -238 (kHz) \\ \hline
	     $\left(E_{\ket{21}}-E_{\ket{11}}\right)-\left(E_{\ket{20}}-E_{\ket{10}}\right)$ & $ZZ_{2110}$ & -148 (kHz) \\ \hline
	     $\left(E_{\ket{12}}-E_{\ket{11}}\right)-\left(E_{\ket{02}}-E_{\ket{01}}\right)$ & $ZZ_{1021}$ & -183 (kHz) \\ \hline
	     $\left(E_{\ket{22}}-E_{\ket{12}}\right)-\left(E_{\ket{20}}-E_{\ket{10}}\right)$ & $ZZ_{2120}$ & -211 (kHz) \\ \hline
	     $\left(E_{\ket{22}}-E_{\ket{21}}\right)-\left(E_{\ket{02}}-E_{\ket{01}}\right)$ & $ZZ_{2021}$ & -262 (kHz) \\ \hline
	     $\left(E_{\ket{12}}-E_{\ket{02}}\right)-\left(E_{\ket{10}}-E_{\ket{00}}\right)$ & $ZZ_{1020}$ & -402 (kHz) \\ \hline
	     $\left(E_{\ket{21}}-E_{\ket{20}}\right)-\left(E_{\ket{01}}-E_{\ket{00}}\right)$ & $ZZ_{2010}$ & -403 (kHz) \\ \hline
	     Coefficient of $(a^{\dagger}a)(b^{\dagger}b)$ & $J_{11}$ & -304.3 (kHz) \\ \hline
	     Coefficient of $(a^{\dagger}a)^2(b^{\dagger}b)$ & $J_{21}$ & 37.8 (kHz) \\ \hline
	     Coefficient of $(a^{\dagger}a)(b^{\dagger}b)^2$ & $J_{12}$ & 23.6 (kHz) \\ \hline
	     Coefficient of $(a^{\dagger}a)^2(b^{\dagger}b)^2$ & $J_{22}$ & 5.4 (kHz) \\ \hline
	\end{tabular}
	\caption{Experimentally obtained various frequency parameters.}
	\label{table:frequency}
\end{table}

\section{Hamiltonian}
\label{app:Hamiltonian}

\begin{table}[t]
    \begin{tabular}{||c|c|c|c||}
		    \hline
		     Capacitance & (fF) & Josephson energy & (GHz) \\  \hline
		    $C_{q1}$ & $178.0$ & $E_{J1}$ & $13.6$ \\  \hline
		    $C_{q2}$ & $131.0$ & $E_{J2}$ & $13.3$ \\  \hline
		    $C_{c}$ & $193.6$ & $E_{Jc}$ & $1140.0$ \\  \hline
		    $C_{q12}$ & $2.0$ &  &  \\  \hline
	\end{tabular}
		\caption{Estimated capacitances and Josephson energies. Capacitances are extracted from Ansys Q3D simulations. Josephson energies are estimated using Ambegaokar-Baratoff relation and room temperature resistances of on-chip (nominally identical) test junctions.}
	\label{table:device_param}
\end{table}

We develop a simplified model to extract various parameters and explain the sideband interactions. Omitting the readout resonators, the bare Hamiltonian can be expressed as
\begin{align}
    H = & \overrightarrow{q}^{\intercal}C_{L}^{-1}\overrightarrow{q} -E_{J1}\cos{\left(\phi_{c}-\phi_{1}\right)}-E_{J2}\cos{\left(\phi_{c}-\phi_{2}\right)} \nonumber, \\ 
    & -E_{Jc}\cos{\left(2\pi\frac{\Phi_{\rm ext}}{\Phi_0}\right)}\cos{\left(\phi_{c}\right)}, \label{eq:H1}
\end{align}
with the capacitance matrix
\begin{equation}
    C_L = \begin{bmatrix}
               C_{1}+C_{12} & -C_{12} & 0 \\
               -C_{12} & C_{2}+C_{12} & 0 \\
               0 & 0 & C_{1}+C_{2}+C_{c}
    \end{bmatrix}.
    \label{eq:H3}
\end{equation}  
Here $\{ \phi_1, \phi_2, \phi_c \}$ are the node fluxes for the two qubits and the coupler respectively. $\overrightarrow{q}^{\intercal} = \left(q_{1}, q_{2}, q_{c}\right)$ represent the charge vector conjugate to the node fluxes, so that $\left[ \phi_{k}, q_k\right] = i$. The coupler capacitance $C_{c}$ originates from the self-capacitance of the large coupler junctions.

\begin{figure}[t]
    \centering
    \includegraphics{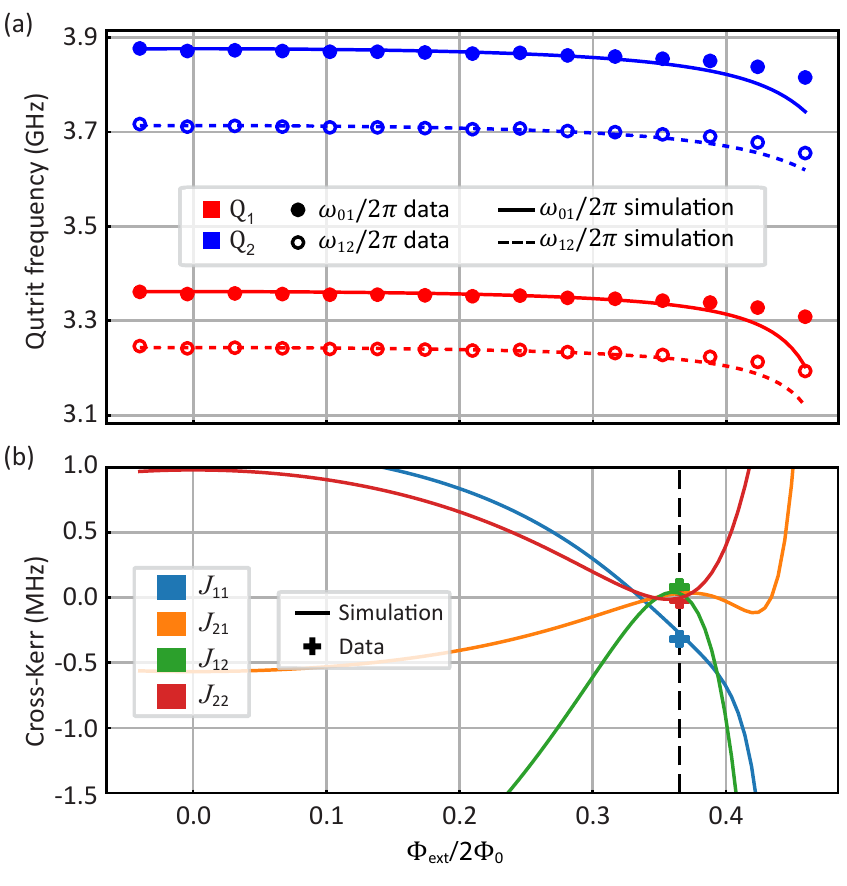}
    \caption{Circuit quantization results. (a) Comparison of the experimentally obtained $\ket{0}\leftrightarrow \ket{1}$ transition frequency $\omega_{01}$ (solid circle) and $\ket{1}\leftrightarrow \ket{2}$ transition frequency $\omega_{12}$ (hollow circle) as a function of $\Phi_{\rm ext}$ with simulations (solid/dashed lines) for qutrits $Q_1$ (red) and $Q_2$ (blue). (b) Different cross-Kerr parameters $J_{11}$ (magenta), $J_{21}$ (orange), $J_{12}$ (green), $J_{22}$ (brown) as a function of $\Phi_{\rm ext}$. The solid lines are obtained from the numerical calculation, and the experimental values at the operating flux-bias (dashed line) are shown with plus symbols.}
    \label{fig:quantization}
\end{figure}

We first extract the linear part of the Hamiltonian,
\begin{multline}
    H_{\rm lin} = \overrightarrow{q}^{\intercal}C_{L}^{-1}\overrightarrow{q}+\frac{E_{J1}}{2}\left(\phi_{c}-\phi_{1}\right)^{2}+\frac{E_{J2}}{2}\left(\phi_{c}-\phi_{2}\right)^{2} \\
    +\frac{E_{Jc}}{2}\cos{\left(2\pi\frac{\Phi_{\rm ext}}{\Phi_0}\right)}\phi_{c}^{2} \label{eq:H2}
\end{multline}    
Table.~\ref{table:device_param} includes all coefficients used in the simulation. We rewrite the charge and phase variables in the dressed basis such that $H_{\rm lin}$ is simultaneously diagonalized for both $\overrightarrow{n}$ and $\overrightarrow{\phi}$ with the transformation matrix $U$:
\begin{subequations}
\begin{align}
     &H_{\rm lin} = \sum_{k=1,2,c}\left(\tilde{C_{k}}\tilde{n}_{k}^{2}+\tilde{D}_{k}\tilde{\phi}_{k}^{2}\right) \\
     &\overrightarrow{\tilde{n}} = \left(\tilde{n}_{1},\tilde{n}_{2},\tilde{n}_{c}\right)^{\intercal} = U^{-1}\overrightarrow{n} \\
     &\overrightarrow{\tilde{\phi}} = \left(\tilde{\phi}_{1},\tilde{\phi}_{2},\tilde{\phi}_{c}\right)^{\intercal} = U^{-1}\overrightarrow{\phi} \\
     & U =
    \begin{bmatrix}
               U_{11} & U_{12} & U_{1c} \\
               U_{21} & U_{22} & U_{2c} \\
               U_{c1} & U_{c2} & U_{cc} 
    \end{bmatrix}.
\end{align}
\end{subequations}
Then we quantize the circuit on the dressed basis,
\begin{subequations}
\begin{align}
     &\tilde{n}_{k} =\frac{i}{\sqrt{2}}\sqrt{\frac{\tilde{D}_{k}}{\tilde{C}_{k}}}\left(a_{k}^{\dagger}-a_k\right), \\
     &\tilde{\phi}_{k} =\frac{1}{\sqrt{2}}\sqrt{\frac{\tilde{C}_k}{\tilde{D}_k}}\left(a_{k}^{\dagger}+a_k\right),
\end{align}
\end{subequations}
and insert the nonlinear part back into the Hamiltonian,
\begin{multline}
H = \sum_{k=1,2,c}\left(\tilde{C}_{k}\tilde{n_{k}}^{2}\right)-E_{J1}\cos{\left(\sum_{k=1,2,c}\left(U_{ck}\tilde{\phi}_{k}-U_{1k}\tilde{\phi}_{k}\right)\right)} \\
-E_{J2}\cos{\left(\sum_{k=1,2,c}\left(U_{2k}\tilde{\phi}_{k}-U_{ck}\tilde{\phi_{k}}\right)\right)} \\ 
-E_{Jc}\cos{\left(2\pi\frac{\Phi_{\rm ext}}{\Phi_0}\right)}\cos{\left(\sum_{k=1,2,c}U_{ck}\tilde{\phi}_{k}\right)}.
\end{multline}
The resulting transition frequencies and various $ZZ$ interactions strengths as a function of $\Phi_{\rm ext}$ obtained from the numerical diagonalization along with the experimental data are shown in Fig.~\ref{fig:quantization}. We notice some deviation when the DC flux is biased very close to $\frac{\pi}{2}$. This is caused by the parasitic SQUID-loop inductance. However, around the DC flux position ($\Phi_{\rm ext}=$0.185$\Phi_0$) where we implement all quantum algorithms (vertical dashed line in Fig.~\ref{fig:quantization}(b)), we have a good match between the numerical and experimental data.

To understand how two-qutrit sidebands work, we follow Ref.~\onlinecite{Yao2017stabilization} and apply adiabatic approximation to the Hamiltonian. The frequency of the coupler mode remains well above ($>15$ GHz) any other frequencies ($<4$ GHz) in the system and thus can be assumed to stay in the ground state during any operation. The static energy of the coupler mode is removed by minimizing the full circuit Hamiltonian. Treating qutrits as Duffing-type oscillators, a toy model with adiabatic approximation can be written as:
\begin{subequations}
\begin{align}
     H_{toy} = & \omega_{1}a_{1}^{\dagger}a_{1}+\omega_{2}a_{2}^{\dagger}a_{2}+\frac{\alpha_{1}}{2}a_{1}^{\dagger}a_{1}^{\dagger}a_{1}a_{1}+\frac{\alpha_{2}}{2}a_{2}^{\dagger}a_{2}^{\dagger}a_{2}a_{2} \nonumber\\
     & +g_{1}\left(t\right)\left(a_{1}^{\dagger}+a_{1}\right)\left(a_{2}^{\dagger}+a_{2}\right) \nonumber\\
     & +g_{2}\left(-a_{1}^{\dagger}+a_{1}\right)\left(-a_{2}^{\dagger}+a_{2}\right), \\
     g_{1}\left(t\right) = & \frac{\sqrt{E_{J1}E_{J2}}}{2E_{Jc}\cos{\left(2\pi{\Phi_{\rm ext}\left(t\right)}/\Phi_0\right)}} \sqrt{\omega_{1}\omega_{2}}, \\
     g_{2} = & \frac{\sqrt{C_{1}C_{2}}}{2C_{12}}\sqrt{\omega_{1}\omega_{2}}.
\end{align}
\end{subequations}
Here $g_{1}\left(t\right)$ and $g_{2}$ are inductive and capacitive coupling strengths respectively. $g_{1}\left(t\right)$ is flux tunable, and when the RF flux modulation frequency matches with any two-photon transition frequency $\ket{jk}\leftrightarrow\ket{j+1,k-1}$ or $\ket{jk}\leftrightarrow\ket{j+1,k+1}$, the corresponding sideband is activated in the system. This adiabatic approximation requires sideband frequencies to be much smaller than qutrit frequencies, which can capture the (red) sideband rates for $\ket{22}\leftrightarrow\ket{31}$ and $\ket{21}\leftrightarrow\ket{30}$ used in the system. However, the (blue) sideband rate for $\ket{jk}\leftrightarrow\ket{j+1,k+1}$ cannot be explained with such an approximation and requires keeping the coupler mode in the analysis~\cite{IBM2017}.

\section{CPhase gates}
\label{app:CPhase}
\begin{figure}[t]
    \centering
    \includegraphics[width=1.0\columnwidth]{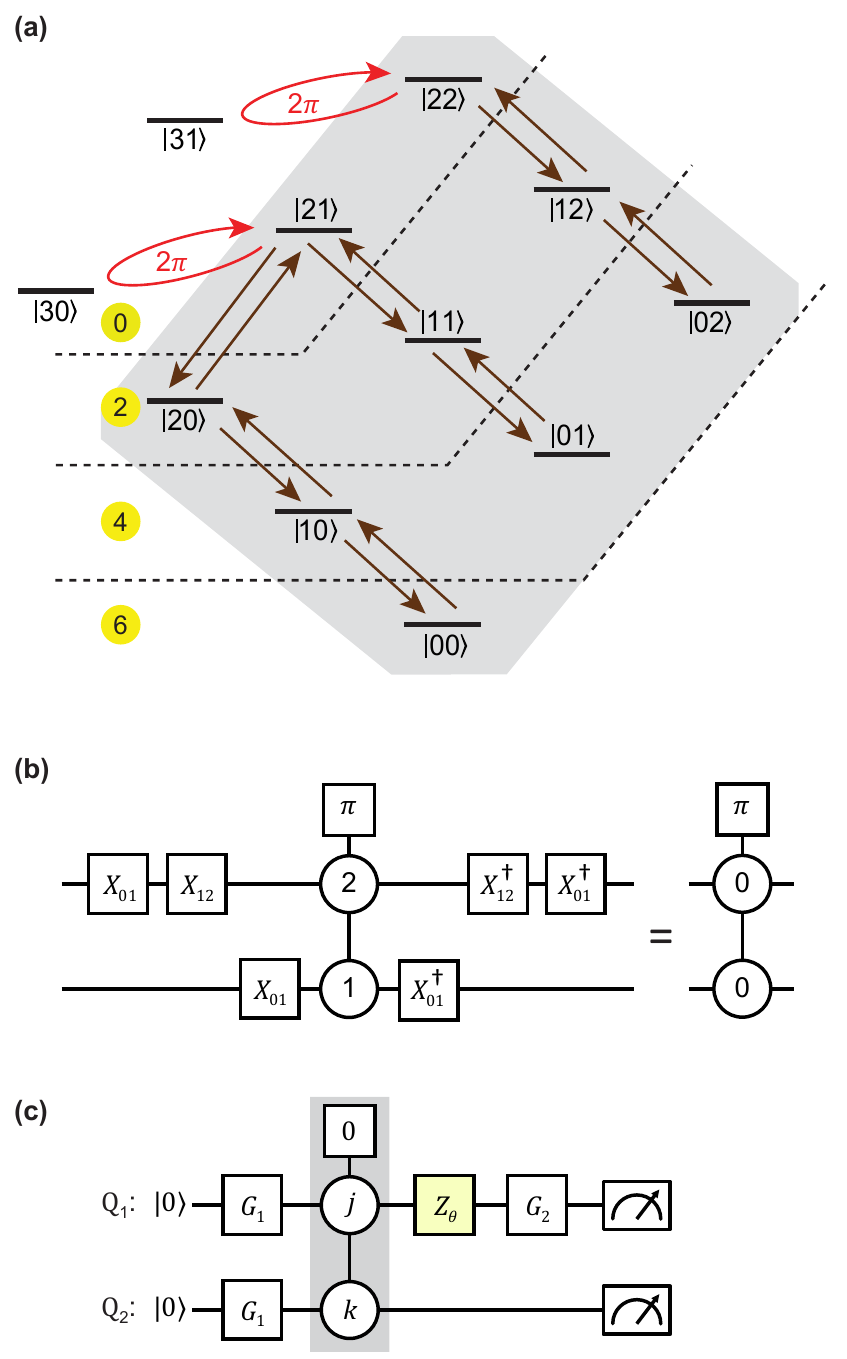}
    \caption{CPhase gate construction. (a) Partitioned energy level diagram. Flipping the phase of a specific target state in region $\left(0\right)$ is performed through a $2\pi$ sideband rotation (shown in red arrow). Flipping the other target states requires a decomposition. The application of a $C_p(\pi,\ket{mn})$ follows the path marked with brown arrows starting from $\ket{mn}$. Here the numbers inside the yellow circle indicate the total number of single-qutrit $\pi$ pulses required. (b) An example of circuit decomposition for the $C_{p}\left(\pi, \ket{00}\right)$ gate. (c) Single-qutrit phase compensation calibration for a $C_{p}\left(\pi, \ket{jk}\right)$ gate.}
    \label{fig:Grover_level}
\end{figure}
As shown in Fig.~\ref{fig:Grover_level}(a), we utilize two native CPhase gates $C_{p}\left(\theta,\ket{21}\right)$ and $C_{p}\left(\theta,\ket{22}\right)$, which can be realized by applying $2\pi$ rotations to $\ket{21} \leftrightarrow \ket{30}$ and $\ket{22} \leftrightarrow \ket{31}$ transitions with a phase difference of $\pi-\theta$ between the first and the last $\pi$ rotations. The optimized gate lengths are $55.9$~ns and $94.0$~ns for $C_{p}\left(\theta,\ket{21}\right)$ and $C_{p}\left(\theta,\ket{22}\right)$ respectively.

The CPhase gates on the states in the region $\left(2\right)$, $\left(4\right)$ and $\left(6\right)$ (highlighted with yellow) need to decomposed where the number indicates the number single-qutrit rotations required. The decomposition starts from the target state, followed by the application of single-qutrit rotations to arrive at one of the states in the region $\left(0\right)$ (following brown arrows). After applying the native CPhase gate, reverse single-qutrit rotations are administered traversing the same path back to the target state. Fig~\ref{fig:Grover_level}(b) shows the circuit decomposition of $C_{p}\left(\pi,\ket{00}\right)$ as an example, which requires the maximum number of pulses. Note that, even though we have access to $C_{p}\left(\pi,\ket{12}\right)$, we do not use it due to poorer fidelity.

\begin{figure}[t]
    \centering
    \includegraphics[width=0.5\textwidth]{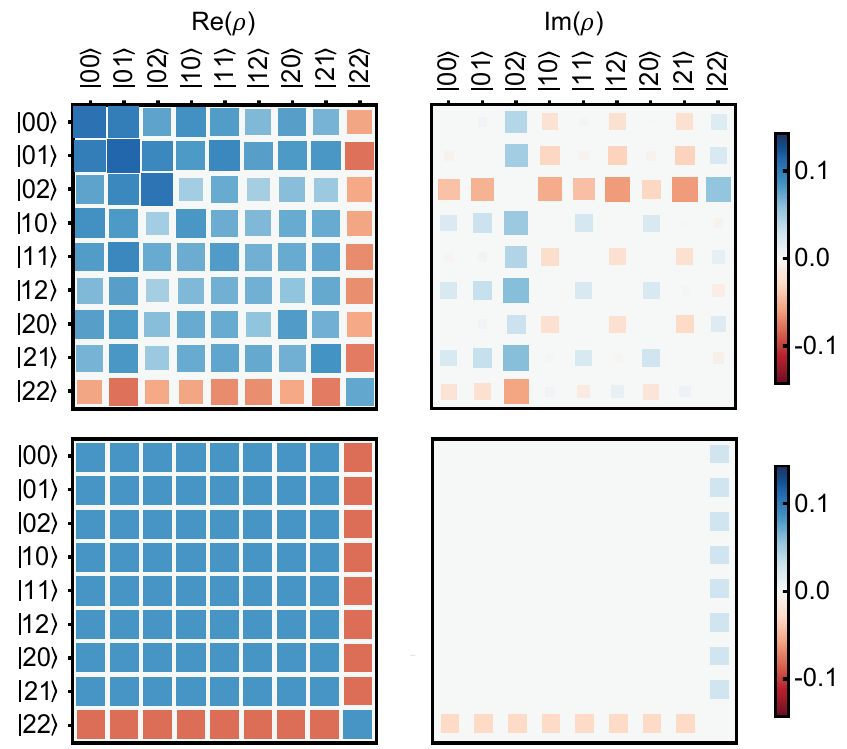}
    \caption{Two-qutrit tomography of the state $\frac{1}{3}\left(\ket{0}+\ket{1}+\ket{2}\right)^{\otimes 2}$ after applying the CPhase gate $C_{p}\left(\frac{8\pi}{9},\ket{22}\right)$. The top and bottom rows are experimental and ideal density matrices, with the real and imaginary parts shown in the left and right columns.}
    \centering
    \label{fig:CPhase}
\end{figure}

The calibration of the CPhase gates involves compensation for the additional phases acquired during the gate operation~\cite{IBMcalib2020}. In our coupler, both the AC stark shift and the rectification effect from SQUID flux modulation can cause qutrit frequencies to change during the gate, resulting in extra phases $\beta_{01}^{(j)}$ and $\beta_{12}^{(j)}$ for the $\ket{0} \leftrightarrow \ket{1}$ and $\ket{1} \leftrightarrow \ket{2}$ transitions of the $j$-th qutrit. We use the circuit shown in Fig.~\ref{fig:Grover_level}(c) to extract these additional phases. Starting from the ground state, we apply the gate $G_{1}=R_{01}\left(0,\frac{\pi}{2}\right)$ to both qutrits if calibrating $\beta_{01}^{(j)}$, followed by $C_{p}\left(0,\ket{jk}\right)$ and sweep the virtual phase $Z_{\theta} = \Theta(\theta, 0)$ before the last gate $G_{2}=R_{01}\left(\pi,\frac{\pi}{2}\right)$ on $Q_{j}$. By fitting the readout on $Q_{j}$ to $C_{0}+C_1\sin{\left(\beta_{01}^{(j)}+\theta\right)}$, one can extract the extra phase $\beta_{01}^{(j)}$ acquired. A similar procedure is used to extract $\beta_{12}^{(j)}$ where we use $G_{1}=R_{12}\left(0,\frac{\pi}{2}\right) \cdot R_{01}\left(0,\pi\right)$, $G_2=R_{12}\left(\pi,\frac{\pi}{2}\right)$, and $Z_{\theta} = \Theta(0,\theta)$. Hence, each CPhase gate has four virtual phases corresponding to two native transitions for each qutrit. As a demonstration, we apply the CPhase gate $C_{p}\left(\frac{8\pi}{9},\ket{22}\right)$ on the state $\frac{1}{3}\left(\ket{0}+\ket{1}+\ket{2}\right)^{\otimes 2}$ (with initial state fidelity of 87.7\%). The final state (shown in Fig.~\ref{fig:CPhase}) obtained after two-qutrit tomography~\cite{Tanay2021tomography} shows a fidelity of 82.4\%, which is limited by the cross-Kerr coupling between qutrits and SPAM error.

\section{Measurement error mitigation}

\label{app:error_mitigation}
\begin{figure}[b]
    \centering
    \includegraphics{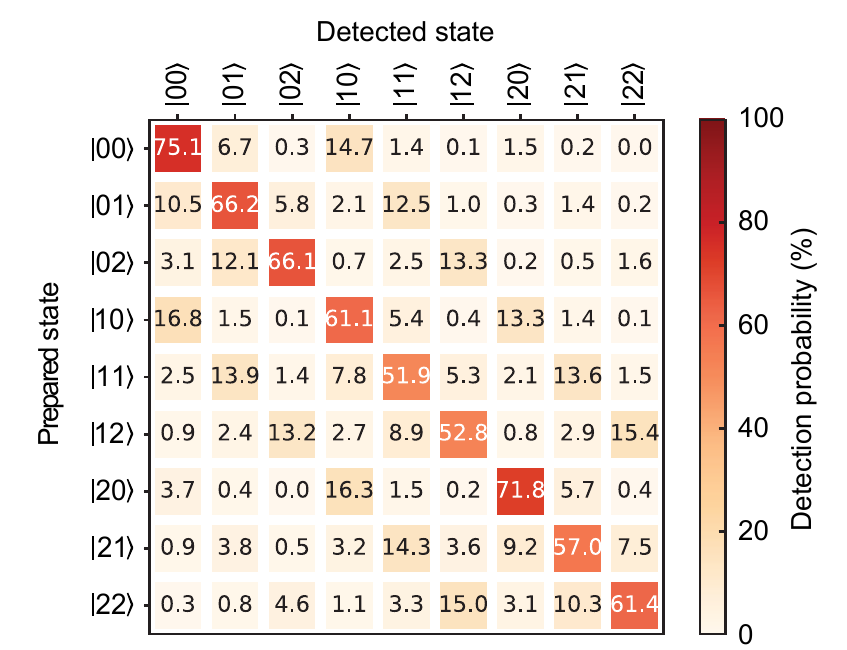}
    \caption{Heatmap of the confusion matrix. $3^2$ basis states are prepared and measured for 20,000 times. The numbers represent average assignment probabilities.}
    \centering
    \label{fig:heatmap}
\end{figure}

Figure~\ref{fig:heatmap} shows the single-shot assignment probability for the nine basis states of our two-qutrit processor. We refer this 2d array as the confusion matrix. Our readout fidelity is limited because there are no parametric amplifiers on the output lines. To fairly demonstrate the performance of the qutrit algorithms with single shot results, we apply the inverse of the confusion matrix to the readout results to compensate for the measurement error. After correcting for the measurement error, it is possible that some of the readout counts (number of times the system found in a specific state) become negative, which happens due to drifts in the calibration parameters. 

This is corrected using Maximum-Likelihood-Estimation (MLE) with the assumption that the minimum fluctuation of a measurement repeated $N$ times should not be lower than $\sqrt{N}$ (assuming normal distribution). We define the following cost function to avoid non-physical measurement counts: 
\begin{equation}
f_c(\overrightarrow{p},\overrightarrow{q})=\sum_{i=1}^{9}\left(\frac{p_{i}-q_{i}}{q_{i}}\right)^2, 
\end{equation}
with the restriction $p_{i}\geq{\sqrt{N}}$. Here $\overrightarrow{q}$ contains the experiment counts (1d array of 9 elements corresponding to the basis states) after applying the inverse of the confusion matrix, and $\overrightarrow{p}$ is the extracted counts after MLE. 

\section{Balanced oracles of DJ algorithm}
\label{app:balanced}
We implement 16 different balanced functions whose equivalent classical functions are tabulated in Table~\ref{table:balanced}. We use $A$ and $B$ to represent the classical ternary values (0,1, and 2) for the two qutrits. The operator $\oplus$ and $\odot$  correspond to addition and multiplication modulo 3 respectively.

\begin{table}[h]
    \begin{tabular}{||c|c||c|c||}
		    \hline
		     Oracle & Classical function & Oracle & Classical function \\  \hline
		    $Z \otimes I$ & A $\oplus$ 0  & $I \otimes Z$ & 0 $\oplus$ B\\  \hline
		    $Z \otimes X$ & A $\oplus$ 1  & $X \otimes Z$ & 1 $\oplus$ B\\  \hline
		    $Z \otimes X^2$ & A $\oplus$ 2  & $X^2 \otimes Z$ & 2 $\oplus$ B\\  \hline
		    $Z^2 \otimes I$ & (2 $\odot$ A) $\oplus$ 0  & $I \otimes Z^2$ & 0 $\oplus$ (2 $\odot$ B)\\  \hline
		    $Z^2 \otimes X$ & (2 $\odot$ A) $\oplus$ 1  & $X \otimes Z^2$ & 1 $\oplus$ (2 $\odot$ B)\\  \hline
		    $Z^2 \otimes X^2$ & (2 $\odot$ A) $\oplus$ 2  & $X^2 \otimes Z^2$ & 2 $\oplus$ (2 $\odot$ B)\\  \hline
		    $Z \otimes Z$ & A $\oplus$ B  & $Z \otimes Z^2$ & A $\oplus$ (2 $\odot$ B)\\  \hline
		    $Z^2 \otimes Z$ & (2 $\odot$ A) $\oplus$ B  & $Z^2 \otimes Z^2$ & (2 $\odot$ A) $\oplus$ (2 $\odot$ B)\\  \hline
	\end{tabular}
		\caption{Equivalent classical functions (ternary valued) for the 16 balanced functions implemented in the DJ algorithm.}
	\label{table:balanced}
\end{table}

\section{Error analysis}

\label{app:error}
\begin{figure}[t]
    \centering
    \includegraphics{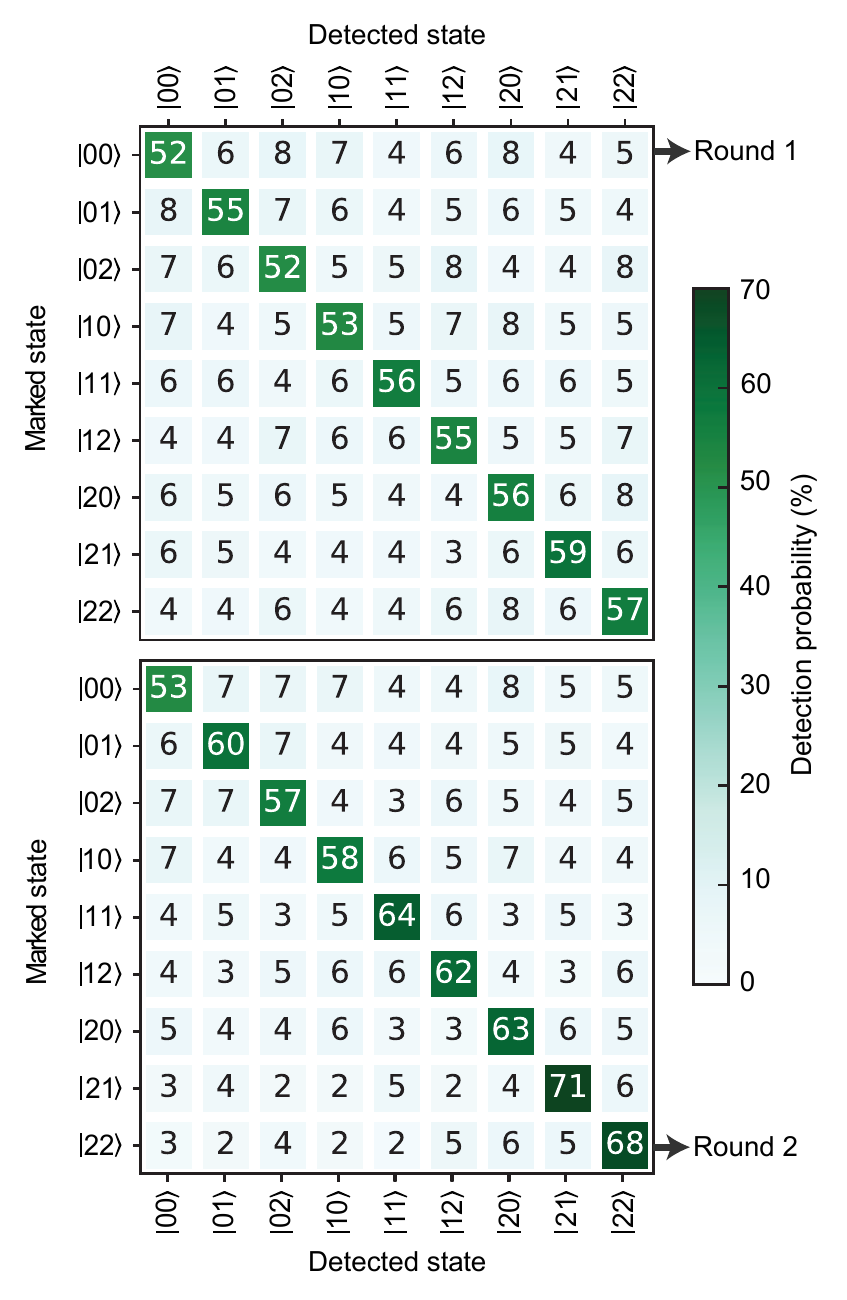}
    \caption{Master equation simulation of two-qutrit Grover's search with one and two stages of amplitude amplification. These results are in a good agreement with the experimental outcomes shown in Fig.~\ref{fig:Grover}(b).}
    \centering
    \label{fig:Grover_simulation}
\end{figure}

Multiple error sources limiting us from approaching the theoretical success rates for different quantum algorithms. The dominant one is the dephasing noise. The deepest circuit (2 stages of Grover's search) implemented includes $17$ single-qutrit and $4$ two-qutrit gates, with a total execution time of $2.11~\mu$s. This time becomes comparable to the qutrit's Ramsey times (see Table~\ref{table:qutrit_coherence}) and strongly affects the success rates. The second source is the inter-qutrit dispersive coupling, including both the static-$ZZ$ values measured in Table.~\ref{table:frequency} and the dynamic-$ZZ$ terms induced during (parametric) gate operations. The $ZZ$ interactions introduce unwanted phase accumulations that reduce the performance of both single-qutrit and CPhase gates. The third source is the leakage to non-computational levels. However, our maximum gate rates are much smaller than the energy gaps, and thus errors due to leakage should not be significant. In order to verify that we have captured all the main sources of error, we perform master-equation simulation using the experimentally obtained device parameters. The simulated results for Grover's search are shown in Fig.~\ref{fig:Grover_simulation}, and those match very well with the averaged square statistical overlap~\cite{Tanay2020Trimon2} of $93.27\%$ and $94.46\%$ respectively for one and two stages of amplitude amplification.

\section{Scaling up}

\begin{figure}[t]
    \centering
    \includegraphics{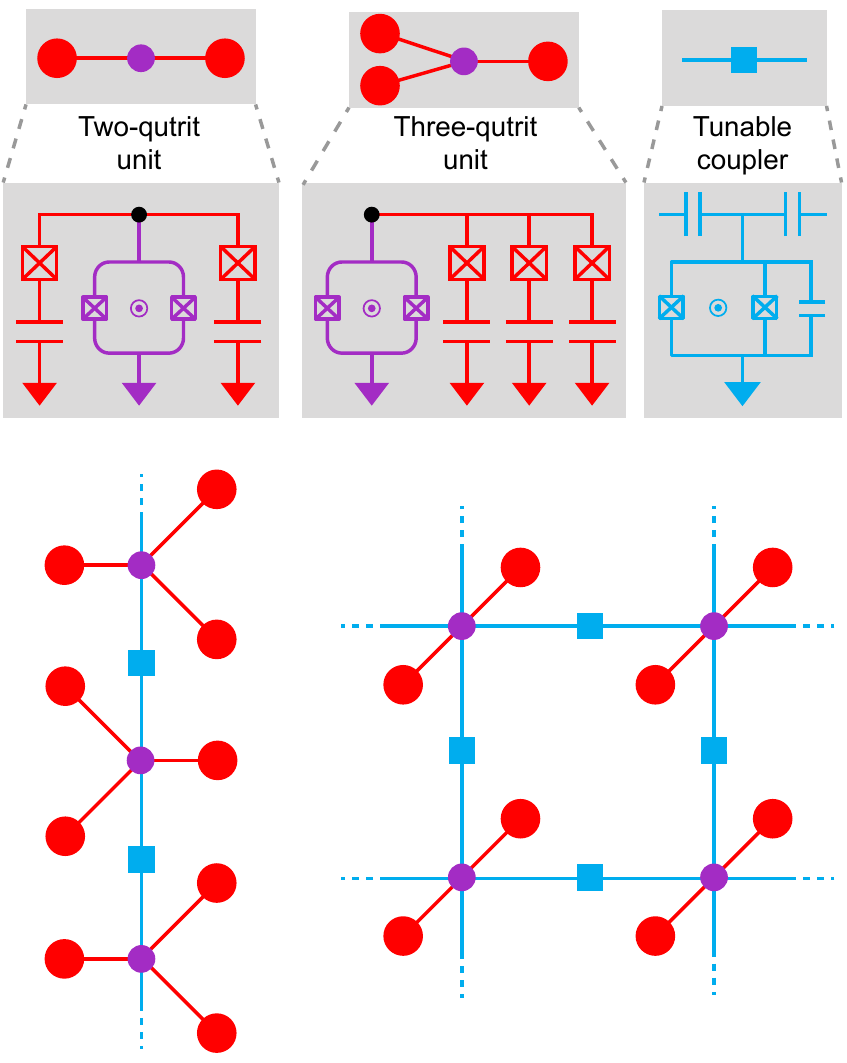}
    \caption{Scaling-up schemes using qutrit-based units as building blocks. }
    \centering
    \label{fig:sacle_up}
\end{figure}

We envisage building a larger system using qutrit-based units, and there are several choices. First, one can couple three transmons to the same parametric coupler~\cite{Ranzani2022stabilization} to build a three-qutrit processor where inter-qutrit static ZZ can be simultaneously minimized through an optimized layout ensuring almost identical inter-transmon capacitances. Next, these units can be used as building blocks to expand in a linear or planar geometry, as shown in Fig.~\ref{fig:sacle_up}. Two units could be coupled via commonly used tunable couplers~\cite{IBM2016tunable_coup, Oliver2020tunable_coup, Rigetti2021tunable_coup} where a galvanic connection is not required. Any two qutrits between two neighboring units can be made equally distant (with respect to hopping) when the capacitive coupling for the inter-unit coupler is made to the central superconducting islands. This way one can realize a moderately-sized qutrit processor consisting of highly efficient two or three qutrit units with fast gates and a slightly slower but switchable interaction between neighboring pairs with almost identical rates.
 
\section{Setup}
\begin{figure*}[t]
    \centering
    \includegraphics{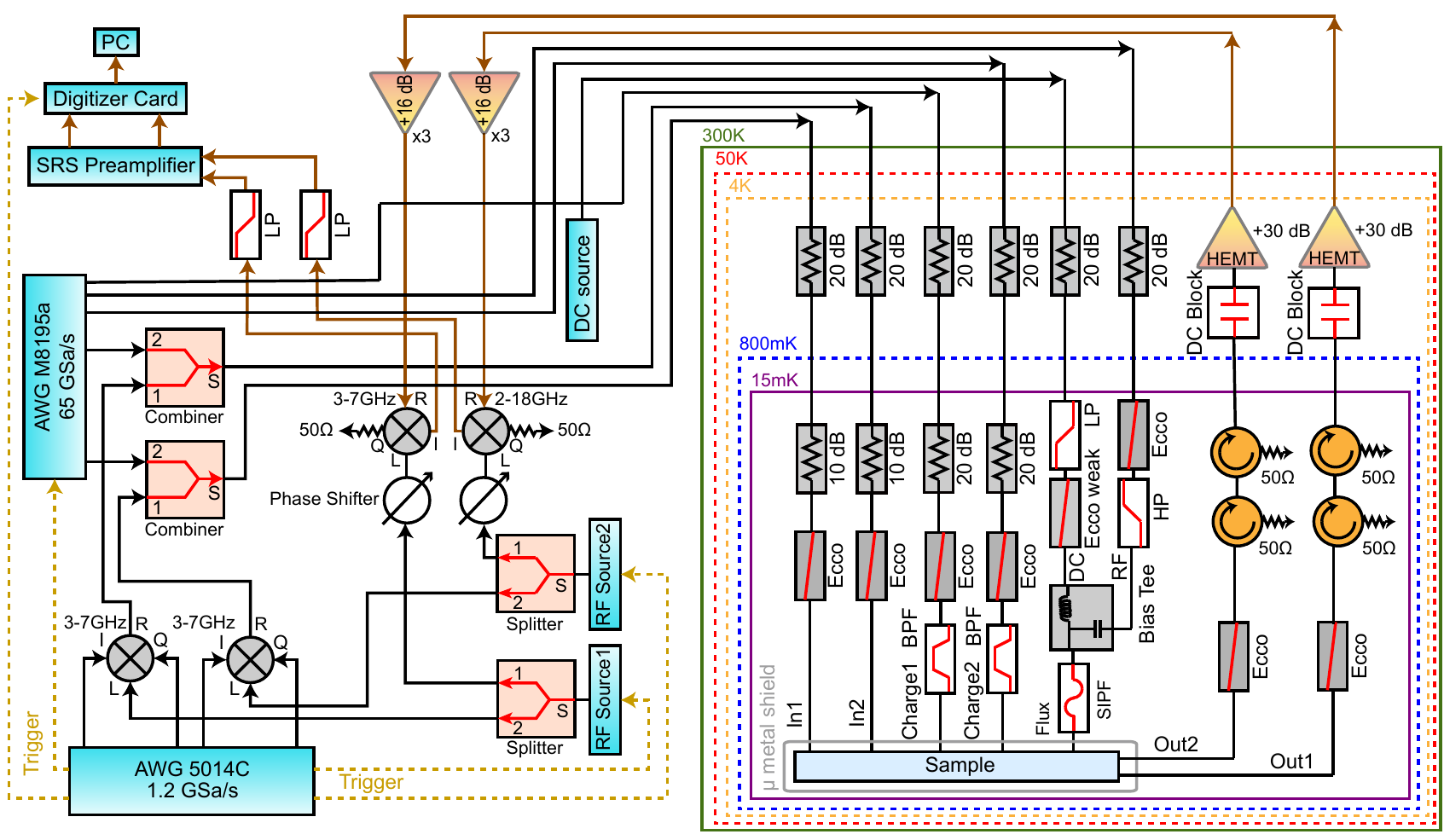}
    \caption{Detailed cryogenic and room-temperature measurement setup.}
    \centering
    \label{fig:measurement}
\end{figure*}

The room and cryogenic temperature wiring diagram is shown in Fig~\ref{fig:measurement}. The device is placed inside a bilayer $\mu$-metal shield, mounted on the base plate of a dilution fridge with 15~mK base temperature. A Tektronix 5014C (1.2GSa/s) Arbitrary Waveform Generator (AWG) acts as the master trigger for all other equipment. The readout pulses are generated through two CW tones from RF sources (PSG-E8257D), modulated by the AWG 5014C. The qutrit pulses are generated through another 4-channel AWG (Keysight M8195 65 GSa/s, 16 GSa/s per channel). Before entering the fridge, the qutrit and readout pulses are combined and sent through lines In$_{1}$ and In$_{2}$. One DC source (Yokogawa GS200) is used to bias the coupler's DC flux position. The RF flux modulation is synthesized through the same 4-channel AWG. Two other direct charge drives, also synthesized through the same AWG, are not used in this project. Inside the fridge, at the 4K plate all input lines have 20 dB attenuation. At the base plate, In$_{1}$ and In$_{2}$ lines have 10-dB attenuation each, followed by a strong Eccosorb\textsuperscript{\textregistered} providing 20 dB attenuation at 4 GHz; Charge$_{1}$ and Charge$_{2}$ lines have 20 dB attenuation, followed by a strong Eccosorb providing 20 dB attenuation at 4 GHz, and a band pass filter with pass band $3.9 - 4.8$~GHz. The DC Flux line has a low pass filter (DC $-~1.9$~MHz), followed by a weak Eccosorb, and the RF Flux line passes through a weak Eccosorb first, followed by a high-pass filter (with a cut-off at $200$~MHz). The DC and RF flux lines are combined with a Bias Tee, and then the signal passes through a Step Impedance Purcell Filter (SIPF), which blocks the frequency band $2.5 - 5.5$~GHz. The output signal goes through a weak Eccosorb, two circulators, a DC block, and is amplified with one LNF HEMT amplifier at 4K. The output signals are further amplified at room temperature and then demodulated with IQ mixers. The demodulated signals pass through low-pass filters (DC $-~250$~MHz) and are amplified again using the SRS Preamplifier. The final signals are digitized with Alazar ATS 9870 (1~GSa/s) and analyzed in a computer.

\normalem{}
\bibliography{algo}
\end{document}